# A Slow Read Attack Using Cloud

Darine Ameyed
Departement of Software
Engineering and IT
École de technologie supérieure ÉTS
Montreal, Canada
darine.ameyed.1@ens.etsmtl.ca

Fehmi Jaafar
School of Computing
Queen's University, Kingston,
Ontario. Canada.
jaafar@cs.queensu.ca

Jaouhar Fattahi
Department of Computer
Sciences and Software
Engineering LSI
Laval University, Quebec,
Canada
jaouhar.fattahi.1@ulaval.ca

*Abstract* – Cloud computing relies on sharing computing resources rather than having local servers or personal devices to handle applications. Nowadays, cloud computing has become one of the fastest growing fields in information technology. However, several new security issues of cloud computing have emerged due to its service delivery models. In this paper, we discuss the case of distributed denial-of-service (DDoS) attack using Cloud resources. First, we show how such attack using a cloud platform could not be detected by previous techniques. Then we present a tricky solution based on the cloud as well.

Keywords- Slow Read attack; distributed denial-of-service; Cloud computing; Security; Virtual IP address;

I. INTRODUCTION

Cloud computing is a model where IT resources (software and / or hardware) are provided as an on-demand service [1]. Currently, Cloud computing is one of the growing areas of information technology with much benefits such as: costs optimization of storages, scalable services and reliable services with an efficient infrastructure, effective tolerance policy for failures and elasticity. However, the adoption of the cloud raises a number of challenges, related in particular to security issues [2].

The Cloud comprises three models of services [3]: Infrastructure as a Service (IaaS), Platform as a Service (PaaS) and Software as a Service (SaaS). All this models have different security issues that need to be considered. In this paper, we are interested in one issue related to Denial of Service attack using the Cloud. Indeed, this attack uses the three models of services on Cloud as follows: (1 a hacker develops a code on the Cloud using the model Software as a Service (SaaS) to launch multiple connections to a same web server. (2 The connection requests will have different IP address because of the use of virtualization in the two other layers of the Cloud model, Infrastructure as a Service (IaaS) and Platform as a Service (PaaS). The Slow Read attack exploits the fact that modern web servers are not limiting the connection duration if there is a data flow between a web server and a web connection from a client, and uses the possibility to delay the TCP connection virtually with zero or a minimal size of data flow by manipulating the TCP receive window size value. Thus, such attack could make unavailable a service or a Web server [4].

One recommendation to protect against such attacks is to set an absolute connection timeout. However, if the timeout is too short, we risk dropping legitimate slow connections. Moreover, if the timeout is too long, we don't get any protection from attacks. Other solutions recommend a timeout value based on the web server connection length statistics, e.g. a timeout slightly greater than median lifetime of connections should satisfy most of the legitimate clients. A major weakness of these approaches is that they cannot differentiate in general slow legitimate connection from attacks.

A set of conventional solutions [5] proposes to monitor the IP addresses, rather than the local traffic volume. Indeed, these approaches recommend filtering out attack traffic at the victim by detecting the set of attack connections using the same IP address. Although that these approaches are easy to be implemented and deployed, they stay not reliable in front of this kind of attack when it is launched from a Cloud environment. Indeed, the Slow Read attack which uses virtual IP addresses has proven to be a troublesome issue for availability of web servers' services. This paper discusses a simple and effective approach which uses the Cloud and IP address provider verification to prohibit the Slow Read attack using virtual IP addresses.

The remainder of the paper is structured as follows:

- First, in section II, we give a short background and we introduce the DDoS attack concepts.
- Second, in section III, we present the Slow Read attack using the Cloud and we analyze a related case study related to it.
- Finally, in Section IV, we present a solution dealing with this attack and we conclude the paper with a discussion of our contributions and we present our future work.

II. RELATED WORK

*A. Classical Network Attacks*

Using networks becomes essential as well as for individuals, for organizations, and businesses to access or offer services or exchange information. However, these platforms still vulnerable and exposed to several attacks. In This section, we describe briefly some types of known networks attacks.

**Threats in Transit:** In a given network, any host is uniquely identified with the physical mac address of its network interface. This interface will be programmed in such a way, that it only receives packets tagged with the unicast address corresponding to the host, the group multicast address of the host and the global multicast address [6]. A threat occurs when an intruder, reprograms the card, so that it accepts packets tagged with the host's address. This process, called wiretapping, is used to intercept communications either to monitor or to interfere and extract information from the network flow [7]. With this kind of attack, encryption is among the counter-measures to resort to.

**TCP Session Hijacking:** Session hijacking, also known as TCP session hijacking, is a method of taking over a Web user session by clandestinely obtaining the session ID and impersonating as the authorized user. Once the user's session ID has been accessed, the attacker can pretense be the legitimate user and do anything the user is authorized to do on the network.

**Man in the Middle Attack:** man in the middle attack (MITM) aims to intercept communications between 2 parties, without either of them knowing that their communication channel has been compromised [8]. The attacker can, not only, read the flow, but also tamper with it.

**Echo-Chargen Attack:** Chargen is a protocol in the TCP/IP stack, used in trials to measure the throughput [9]. This protocol makes use of both UDP and TCP ports 19. Once a client opens a TCP connection on port 19, the server replies back with a random-character stream until the connection is closed. And each time a host sends UDP datagram on port 19, the server replies back with a random message ranging between 0 to 512 characters. UDP CHARGEN has been frequently used in Denial-of-Service attacks. By spoofing the source addresses, an attacker can bring several hosts to target one or several others machines, thereby overloading both the network and the targeted hosts.

**Smurf Attack:** Smurf attack is a kind of Denial-of-Service (DoS) attack. To launch the Smurf attack attacker send a spoofed Echo-Request message to a network's broadcast IP address [9]. The spoofed Echo-Request message has the victim's IP address as the source IP address. Later, each host receiving the broadcast Echo-Request message will direct an Echo-Reply message to the victim. That which results, overwhelming the victim with a flood of Echo-Reply messages.

**Traffic Redirection:** This attack requires understanding of network routing. Routers use complex algorithms to decide how to route the traffic. The best route usually taking into account a combination of several factors: distance, delay, cost, quality, etc. Each router advises its neighbours on how they can reach other addresses on the network. If a router can be brought to pretend it has the best router, all traffic in the network will converge on a single point, which might bring down the network or deteriorate the traffic quality.

**Attacks on Domain Name Service (DNS):** Domain names might be exploited by means of their administrative procedures, targeting the DNS infrastructure. Attackers aim to make access to a service impossible or very difficult, by overloading key nodes in the DNS infrastructure.

**Distributed Denial of Service (DDoS) Attacks:** A DDoS attack aims to make a server or a service inaccessible by overloading its bandwidth or resources. A DDoS attack consists of multiple simultaneous queries sent from multiple sources in the internet to a single target, thus making it unstable or in the worst case, unavailable.

**Syn Flood Attack:** The SYN flood is another form of DDoS attack. It uses the TCP protocol and consists in sending several SYN queries to the target. The attacker can send several SYN to the target and never acknowledges its replies with an ACK thus overloading its SYN_RECV suffer.

*B. Denial of Service*

Denial-of-service attack is a security threat that can make a machine or a network resource unavailable to users [10]. This attack includes some symptoms such as an unusually slow network performance or an inability to access to a web site.

The first method used by hackers to execute a Denial-of-Service attack was to send some malformed packets to the victim to confuse a protocol or an application running on it (i.e., vulnerability attack [11]). The most common methods were to disrupt a legitimate user's connectivity in the network/transport-level [11] or to disrupt a legitimate user's services by exhausting the server resources (e.g., memory, disk, input-output bandwidth) in the application-level [12].

We notice two challenges for detecting Denial-of-Service attacks. The first one is the difficulty to detect such attack when they are highly distributed, because the attack traffic from each source could be smaller than the normal legitimate traffic. The second challenge is to detect such attacks without creating false alarm. Previous approaches were based on monitoring the number of new source IP addresses or the local traffic volume to detect abnormal traffic [13-14].

In this paper, we are interested in one specific method of the Denial-of-Service attack: the Slow Read attack [15]. Indeed, in this type of attack, the hacker sends legitimate application layer requests but reads responses very slowly, thus trying to exhaust the server's connection pool. The hacker advertises a small number for the TCP Receive Window size to ensure a low data flow rate (Figure1). If a hacker sends numerous legitimate requests, the Web server will eventually reach its maximum capacity and becomes unavailable for new connections. Moreover, it is impossible to detect these attacks if they come from the Cloud by monitoring the network layer, because those requests are indistinguishable from other legitimate clients and have different virtual IP address.

*C. Security and Cloud Computing*

Cloud Computing offers dynamically scalable resources provisioned as a service over the Internet. There are three major security concerns that hinder the adoption of cloud services [16]. The first concern is the loss of control over sensitive data [17]. Indeed, current control measures do not adequately address cloud computing third-party data storage and processing needs.

The second concern is the lack of isolation. Indeed, each cloud client may provide one or more services; each of them in turn may have a multitude of users. Thus, it is necessary to ensure that such services can guarantee that their users' data remain private and secured [18]. However, it is hard to achieve that for applications belonging to different cloud tenants or isolate them to prevent data leakage.

The last concern is the capability of using virtualization to hide classical attack scenarios such as the Slow Read attack for Denial-of-service. Indeed, cloud applications are by design opened to the internet, which makes them either target to cyber-attacks such as Denial of Service (DoS) or a platform used to make such attacks [2].

The architectural design and characteristics of cloud computing (based on the centralization of security functionalities, the data and process segmentation, and the high the availability) proposes a number of security benefits [19]. However, the different models of service proposed on the Cloud bring with them several security challenges [20].  These security challenges are mainly related to three concepts [21] [22] : the confidentiality (a set of measures undertaken to prevent sensitive information from reaching the wrong people, which can limits access to data but ensuring  that the legitimate users can get it), the integrity (maintaining and assuring the accuracy and consistency of data on the Cloud over its entire life-cycle. This means that data cannot be modified in an unauthorized or undetected manner), and the availability (remaining systems and data available at all times and preventing service disruptions due to different issues such as hardware failures, software crashes, and  denial-of-service attacks).

Figure 1. The Slow Read attack

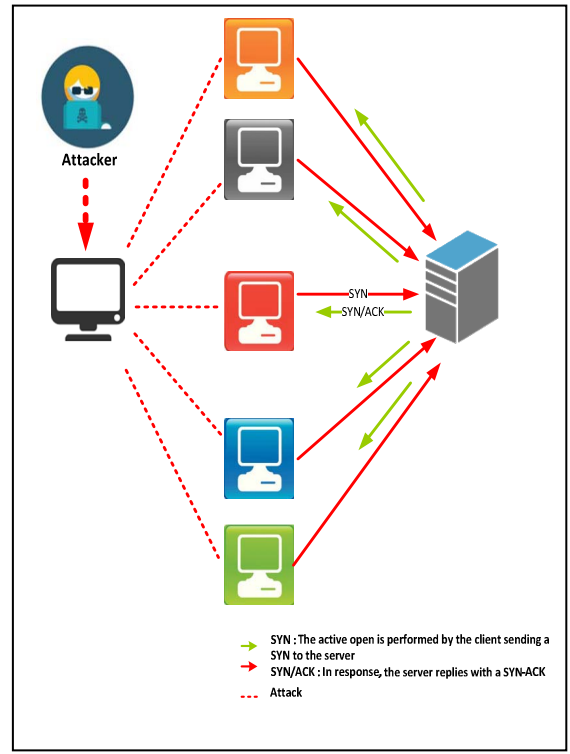

III.  THE SLOW READ ATTACK USING CLOUD

*A. The Case study*

Figure 2 shows the scenario of a Slow Read attack for Denial of Services using a Cloud environment. In this scenario, the attacker sends a set of HTTP requests to a Web server using a program installed on a virtual machine on the cloud. Indeed, each HTTP request has a different IP address although they were launched from the same virtual machine. This is due to the principle of virtualization and resource sharing using Cloud environment. Then, the attacker sets the window size to the smallest acceptable value to make the HTTP response operation slow down. If the number reaches the limit of maximum clients of the web server, the web server cannot accept new legitimate connections and cannot disconnect actual connections. Thus, it becomes inactive and unavailable.

From [23], we deduce three factors that can determine the Slow Read attack.

1. The value of Timeout value of the Web server.

2. The total number of connections that a Web server can process

3. The total number of attack connections from attackers.

To validate this observation, we setup an experiment which consists of executing a program from a virtual machine to launch multiple HTTP requests to a Web server with a receive window range between 8 and 16 bytes and a read rate from receive buffers equal to 5 byte/sec to make the HTTP response operation slow down. From a static point of view, the virtual machine contains a script to execute a Slow Read attack. A second major part is the Web server which runs on a machine separated from the Cloud. We set the number of Max Clients of the web server to 500 and the Timeout for probe connection to 20 seconds. We found that the attack succeeds when the number of connections reaches the limit of Max Clients of the Web server and the system cannot detect such attack because different connections from the same virtual machines have different IP addresses.

Indeed, we observe that if the Timeout is set short, the Slow Read attack can be prevented, but the quality of service is reduced remarkably because legitimate connections could be wrongly disconnected after a short Timeout. If the total number of connections that a Web server can process is set large, we cannot prevent the Slow Read Attack on cloud because a hacker can launch a huge number of connections from the Cloud environment using a simple program knowing that there are always a limitation of resource of Web servers. For the total number of attack connections, the attack connections from the same IP address is easy to be detected. However, using a cloud environment, a hacker can launch a huge number of connections from the same virtual machine but having different IP address. Indeed, the virtualization consists of the masking of server resources, including the number and identity of individual physical servers, processors, and operating systems, from server users. Thus, the server administrator uses a software to divide one physical server into multiple isolated virtual environments. By consequence, a hacker can launch several connection requests from the same virtual machine and using different virtual IP address (an IP address assigned to multiple domain names or servers that share an IP address). Virtual IP addresses are allocated to virtual private servers, websites or any other application residing on a single server. Thus, the classical approaches cannot outface Slow Read attack launched using a Cloud environment.

### B. A Solution Proposal

Figure 3 shows our proposed solution. Indeed, we suggest using Cloud service architecture to assess the availability of web servers' services and outface Slow Read attack. In fact, Cloud provides

Figure 2. A Slow Read attack using a Cloud environment

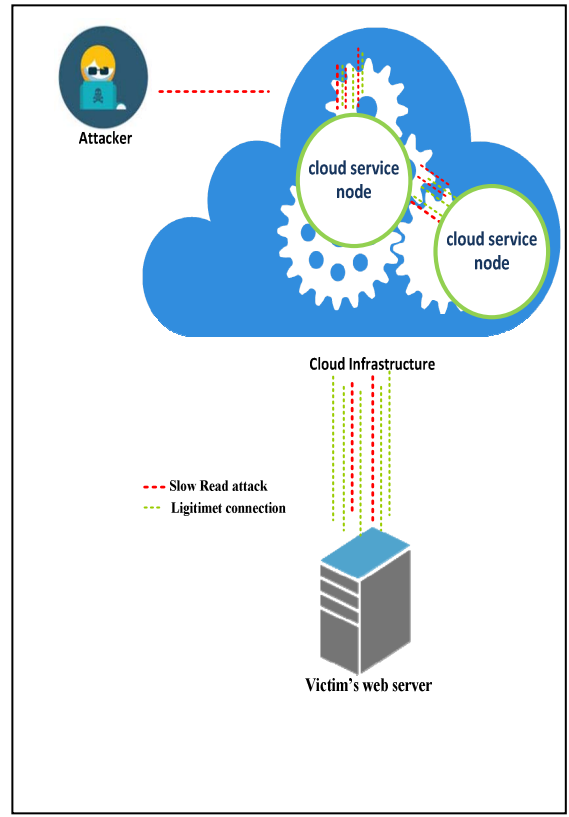

elastic and highly-available computing resources based on server virtualization technology.

We suggest incorporating multiple Web servers by implementing a failure-isolation zone to provide high availability and redundancy in a Cloud environment [24]. This improves performance because several servers can answer user requests simultaneously depending on the traffic. Moreover, web server administrators have to implement such failure isolation zone to protect the environment for unauthorized activity. This provides each node with better security isolation and each web server will be protected from the Slow Read attacks by the availability of one of the two nodes. Indeed, failures in one node do not propagate to the other node and thus, hypervisor provides the visibility of the web server with a more secure and robust environment.

Concretely, in the case of a Slow Read attack, we propose to use a failure-isolation zone to make a high-availability configuration by distributing the web servers' services instances among two zones. In the first zone, when the number of connections reaches the total number of connections that the Web server can process, our approach distributes the new connections to the second web server processes implemented in another zone. Then the approach analyzes the different IP addresses of the actual connection on the first zone to delete slow connections coming from the same IP address or from the same cloud provider as shown in Figure 3 (analysis process). Indeed, it will be considered as suspect the fact to have several slow connections for the same web server coming from the same cloud provider because it may present Slow Read Attack launched using a Cloud environment.

Figure 3. The proposed solution based on Cloud and IP address verification

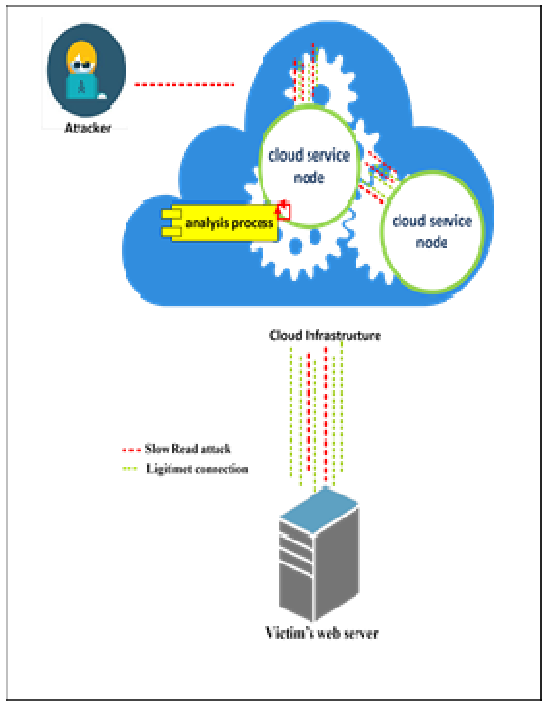

## IV. CONCLUSION

Cloud computing have several architecture based on the services they provide. Cloud's users have to trust the provider on the availability as well as data security. However, several unchartered risks and challenges have been introduced from this relocation to the clouds, deteriorating much of the effectiveness of traditional protection mechanisms. For example, using a cloud environment, a hacker can launch a huge number of Slow Read connections with different virtual IP address to make a Denial of Service attack.

This paper has presented an approach to prohibit the Slow Read attack which uses cloud environment. The solution based on the proposed approach is illustrated with a descriptive web server's service implemented on an IaaS cloud. Our approach is under implementation as a part of system design and security management tools. A future work includes the use of the composed availability model to evaluate the effectiveness of our approach and to measure the ability of our approach to detect and to block Slow Read Attacks launched from a Cloud environment.

REFERENCES

[1] Hogan (M.) et al. – Nist cloud computing standards roadmap, version 1.0. 2011.

[2] Dillon, Tharam, Chen Wu, and Elizabeth Chang. "Cloud computing: issues and challenges." In Advanced Information Networking and Applications (AINA), 2010 24th IEEE International Conference on, pp. 27-33. Ieee, 2010.

[3] Savolainen, Eeva. "Cloud service models." In em Seminar--Cloud Computing and Web Services, UNIVERSITY OF HELSINKI, Department of Computer Science, Helsinki, vol. 10, p. 1012. 2012.

[4] Garber, Lee. "Denial-of-service attacks rip the Internet." Computer 33, no. 4 (2000): 12-17.

[5] Zargar, Saman Taghavi, James Joshi, and David Tipper. "A survey of defense mechanisms against distributed denial of service (DDoS) flooding attacks." Communications Surveys & Tutorials, IEEE 15, no. 4 (2013): 2046-2069.

[6] K. J. Connolly, Law of Internet Security and Privacy, Aspen Publishers, Nov. 2003, ISBN: 0735542732.

[7] C. Pfleeger, Security in Computing, 4th Edition, Prentice Hall, Nov. 2006, ISBN: 0132390779

[8] D. Kaminsky, et. al., Hack Proofing Your Network, Syngress, 2nd Edition, Mar. 2002. ISBN: 1928994709.

[9] J. Postel, Character Generator Protocol, IETF RFC 864, May 1983.

[10] Needham, Roger M. "Denial of service." In Proceedings of the 1st ACM Conference on Computer and Communications Security, pp. 151-153. ACM, 1993.

[11] J. Mirkovic and P. Reiher, A taxonomy of DDoS attack and DDoS defense mechanisms, ACM SIGCOMM Computer Communications Review, vol. 34, no. 2, pp. 39-53, April 2004.

[12] S. Ranjan, R. Swaminathan, M. Uysal, and E. Knightly, DDoS-Resilient Scheduling to Counter Application Layer Attacks under Imperfect Detection, IEEE INFOCOM'06, 2006.

[13] Jaeyeon Jung, Balachander Krishnamurthy, and Michael Rabinovich. Flash crowds and denial of service attacks: Characterization and implications for CDNs and web sites. Proceeding of 11th World Wide Web Conference, May 2002. Honolulu, Hawaii, USA.

[14] Ratul Mahajan, Steven M. Bellovin, Sally Floyd, John Ioannidis, Vern Paxson, and Scott Shenker. Controlling high bandwidth aggregates in the network. Technical report, AT&T Center for Internet Research at ICSI (ACIRI) and AT&T Labs Research, February 2001.

[15] Park, Junhan, et al. "Analysis of Slow Read DoS Attack and Countermeasures." Proceeding of the International Conference on Cyber-Crime Investigation and Cyber Security. 2014.

[16] Intel IT Center, "What's Holding Back the Cloud? Intel Survey on Increasing IT Professionals' Confidence in Cloud Security," May 2012. http://www.intel.com/content/www/us/en/cloud-computing/whats-holding-back-the-cloud-peer-research-report.html

[17] Mehrotra, Sharad. "Towards a Risk-Based Approach to Achieving Data Confidentiality in Cloud Computing." In Secure Data Management, pp. 42-47. Springer International Publishing, 2014.

[18] Krebs, Rouven, Christof Momm, and Samuel Kounev. "Metrics and techniques for quantifying performance isolation in cloud environments." Science of Computer Programming 90 (2014): 116-134.

[19] Bouayad, Anas, Asmae Blilat, N. El Houda Mejhed, and Mohammed El Ghazi. "Cloud computing: security challenges." In Information Science and Technology (CIST), 2012 Colloquium in, pp. 26-31. IEEE, 2012.

[20] Ren, Kui, Cong Wang, and Qian Wang. "Security challenges for the public cloud." IEEE Internet Computing 1 (2012): 69-73.

[21] SO, Kuyoro. "Cloud computing security issues and challenges." International Journal of Computer Networks 3, no. 5 (2011).

[22] Srinivasan, Madhan Kumar, K. Sarukesi, Paul Rodrigues, M. Sai Manoj, and P. Revathy. "State-of-the-art cloud computing security taxonomies: a classification of security challenges in the present cloud computing environment." In Proceedings of the international conference on advances in computing, communications and informatics, pp. 470-476. ACM, 2012.

[23] Park, Junhan, Keisuke Iwai, Hidema Tanaka, and Takakazu Kurokawa. "Analysis of Slow Read DoS Attack and Countermeasures." In Proceeding of the International Conference on Cyber-Crime Investigation and Cyber Security, pp. 37-49. 2014.

[24] Hwang, Kai, and Deyi Li. "Trusted cloud computing with secure resources and data coloring." Internet Computing, IEEE 14, no. 5 (2010): 14-22.